# Time-varying Rotational Inverted Pendulum Control using Fuzzy Approach

R. Eini, S. Abdelwahed

*Abstract*—In this paper, a nonlinear rotational inverted pendulum with time-varying parameters is controlled using the indirect adaptive fuzzy controller design. This type of controller is chosen because this particular system performance is highly sensitive to unavoidable unknown model changes. So, a conventional controller is firstly designed through feedback linearization method, and applied to the system. Feedback linearization method here is used for two purposes; to attain an approximation of necessary system dynamics and to assess the performance of the proposed adaptive fuzzy controller by comparing the results of both adaptive fuzzy and feedback linearization controllers. An indirect adaptive fuzzy controller, resistant to parameter variations is then proposed. The general structure of the adaptive controller is specified in the first stage. In the second stage, its parameters are regulated with the aid of two fuzzy systems. Lyapunov stability theorem is used to regulate the system parameters such that the closed loop system is stabilized and zero tracking error is attained. Finally, the results of the proposed and the conventional approaches are compared. Results showed that the adaptive fuzzy controller performed more efficiently than the classical controller, with existing parameters variations.

*Keywords—feedback linearization; rotational inverted pendulum; indirect adaptive fuzzy controller.*

## I. INTRODUCTION

Fuzzy sets theorem was first introduced by L. A. Zade in 1965 [1]. Unlike classical theory, fuzzy theory is not associated with certain, clear phenomena, so it can be employed successfully in real systems that are usually uncertain and complex. The basis of fuzzy systems when implemented as fuzzy controllers is intelligent decision making based on previous experiences without relying on system modeling. Classical controllers are designed generally based on a mathematical model of the system, whereas fuzzy controllers are designed based on if-then fuzzy rules attained from human operators' experiences or specific knowledge about the system. These rules describe the plant dynamic behaviors and the fuzzy controller is designed by combining these fuzzy logic rules. In many practical control problems, finding an exact mathematical model which is simply perceivable is a difficult task. Therefore, fuzzy systems are widely applicable in different areas of science and engineering applications where the plants are poorly understood and there is a need to approximate true plant representation by fuzzy logic.

Fuzzy controllers are mainly categorized into two types; non-adaptive and adaptive fuzzy controllers. The prominent difference between these two types of fuzzy controllers is the structure of their parameters. Parameters of a non-adaptive fuzzy system are fixed, however these parameters in an adaptive fuzzy system are being updated based on an adaption law. Applying adaptive fuzzy controllers is considered an efficient approach in a wide range of systems since plenty of systems are associated with uncertainties of their structure and their parameters. Adaptive fuzzy controllers are advantageous in terms of performance and efficiency as their parameters can be tuned by environmental changes. Moreover, not too much information of the system being controlled is required because of the existence of identification mechanisms in the adaptive control approach [1], [2].

The first adaptive fuzzy controller was a Linguistic Self-Organizing Controller [3] with many benefits. Fuzzy Model Reference Learning Controller is also introduced as an adaptive fuzzy controller in literatures [4-5] and studies showed effectiveness and applicability of this approach [4-7]. The main problem of applying Linguistic Self-Organizing Controller or Fuzzy Model Reference Learning Controller is that the stability analysis of closed loop systems using these controllers is difficult. Thus, new adaptive fuzzy control design methods are recently being considered for stability analysis of closed loop systems since system stability analysis through these approaches is simple and systematic. Adaptive fuzzy control approaches are becoming more popular because of their extensive applicability in various engineering fields. A number of adaptive fuzzy control techniques can be found in literatures [8-12]. Basically, adaptive fuzzy controllers are categorized into three main groups; indirect adaptive fuzzy controllers, direct adaptive fuzzy controllers, and direct/indirect adaptive fuzzy controllers.

In designing an indirect adaptive fuzzy controller, the fuzzy system is initially constructed based on the plant specific knowledge. Knowledge of the plant gives some information of the system behavior toward various inputs. In the next step, fuzzy controller structure is determined and its parameters are tuned based on the adaption rules such that

R. Eini is Electrical Engineering Department PhD student, Virginia Commonwealth University, Richmond, VA 23220 USA (corresponding author, e-mail: einir@ vcu.edu).

S. Abdelwaed is Electrical Engineering Department faculty, Virginia Commonwealth University, Richmond, VA 23220 USA (e-mail: sabdelwahed@vcu.edu).

not only the closed loop system stability is guaranteed, but also tracking error converges to zero.

The objective of this paper is to design an indirect fuzzy adaptive controller for a nonlinear system with variable parameters and to analyze its superiorities upon a classic controller. In this regard, an indirect adaptive fuzzy control technique is proposed for the first time to control a nonlinear rotational inverted pendulum system. This type of controller has never been designed and implemented for this specific plant before. Using adaptive fuzzy controller for this plant seems reasonable regarding the nonlinear nature of the system with four state variables and unknown time varying parameters. In the first design stage, a classic feedback linearization controller is designed. Next, the adaptive fuzzy controller structure is designed and its parameters are designed using two fuzzy systems. Indeed, adaption mechanisms for the fuzzy system parameters are established based on Lyapunov stability criterion and tracking error minimization. In the final stage, simulation results of applying both classical and proposed controllers are compared and analyzed in detail. Thus, the effectiveness and excellence of the proposed controller in existence of system parametrical uncertainty is proved.

The paper is organized as follows. First, the system formulation and its parameters are represented in detail in section II. Section III and IV describe the feedback linearization control method and proposed adaptive fuzzy control technique for this specific system, and also how to apply them for this plant. Simulation results, analysis and comparison of the techniques are shown in section V. Finally, the paper is concluded in section VI.

## II. System Description and Mathematical Modeling

A nonlinear rotational inverted pendulum system is an important benchmark control problem in dynamics and control theory. This kind of system also has a number of practical applications such as missile launchers, pendubots, segways, earthquake resistant building design, etc. The rotational inverted pendulum system considered in this paper has four state variables [13]; $\theta_0$, $\theta_1$, $\dot{\theta}_0$ and $\dot{\theta}_1$, whereas $\theta_1$ is pendulum angle with respect to the vertical axis, and $\theta_0$ is the rotating base angle of the pendulum (Fig. 1).

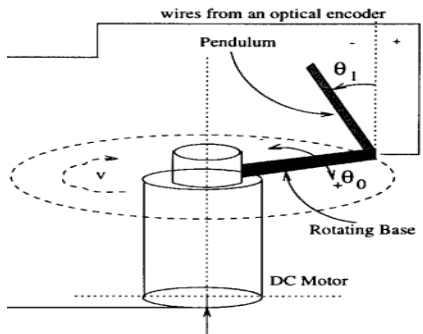

Fig. 1. Rotational inverted pendulum system [14].

Among all state variables, only $\theta_0$ and $\theta_1$ can be measured directly. The two other states ($\dot{\theta}_0$ and $\dot{\theta}_1$) need to be attained with a particular method. Through first-order backward difference approximation, all the states are obtained [15].

Therefore, descriptive formulations of system states are as follows:

$$\ddot{\theta}_0 = a_p \dot{\theta}_0 + k_p u \qquad (1)$$
$$\ddot{\theta}_1 = -\frac{c1}{J1}\dot{\theta}_1 + \frac{m_1 g l_1}{J_1} \sin \theta_1 + \frac{k_1}{J_1}\ddot{\theta}_0$$

In the above equations, $u$ is the system input, $a_p$ and $k_p$ are parameters of dc motor with torque constant $k_1$, $g$ is the acceleration due to gravity, $m_1$ is the pendulum mass, $l_1$ is the pendulum length, $J_1$ is the pendulum inertia, and $c_1$ is a constant associated with friction. The system parameter values are as follows [15]:

$$m_1 = 8.6184 \times 10^{-2} kg, \quad k_1 = 1.9 \times 10^{-3}$$
$$a_p = 33.04, \quad J_1 = 1.031 \times 10^{-3}, \quad g = 9.8066 \qquad (2)$$
$$l_1 = 0.113m, \quad c_1 = 2.979 \times 10^{-3}, \quad k_p = 74.89$$

Considering $\dot{\theta}_1 = x_4$, $\theta_1 = x_3$, $\dot{\theta}_0 = x_2$ and $\theta_0 = x_1$, the following state space equations for the pendulum system is attained.

$$\begin{cases} \dot{x}_1 = x_2 \\ \dot{x}_2 = a_1 x_2 + b_1 u \\ \dot{x}_3 = x_4 \\ \dot{x}_4 = a_2 x_2 + a_3 \sin x_3 + a_4 x_4 + b_2 u \end{cases} \qquad (3)$$

State $x_3$ represents the system output, which is the pendulum angle with respect to the vertical axis. The parameters in (3) are described as follows:

$$a_1 = -a_p, \quad a_2 = -\left(\frac{k_1 a_p}{J_1}\right), \quad a_3 = \frac{m_1 g l_1}{J_1}, \qquad (4)$$
$$a_4 = -\left(\frac{C_1}{J_1}\right), \quad b_1 = k_p, \quad b_2 = \frac{k_1 k_p}{J_1}$$

## III. Feedback Linearization Controller Design

To design a feedback linearization control, relative degree of the system is required [16-17]. By definition, the relative degree is the number of times we have to differentiate the output y before the input u appears explicitly. Thus, by

differentiating state variable $x_3$ which is the system output, the relative degree is 2 (as follows):

$$\dot{y} = \dot{x}_3 = x_4 \Rightarrow \ddot{y} = \dot{x}_4 = a_2 x_2 + a_3 \sin x_3 + a_4 x_4 + b_2 u$$

According to literatures [18], feedback linearization control law for this plant is stated as (5).

$$u_f = \frac{1}{b_2}[-(a_2 x_2 + a_3 \sin x_3 + a_4 x_4) + v] \quad (5)$$

By applying control signal $u_f$, the state space equation (3) is converted to a new linearized equation. Assuming new state variables as $Z = T(x)$, the updated linear state space equation is represented as (6).

$$\begin{cases} z_{11} = T_1(x) = x_3 \\ z_{12} = T_2(x) = x_4 \\ z_{21} = T_3(x) = x_1 + x_3 \\ z_{22} = T_4(x) = x_2 - \left(\frac{b_1}{b_2}\right) x_4 \\ y = z_{11} \end{cases} \quad (6)$$

Furthermore, the zero dynamics of the system are stated in (7) [18].

$$\begin{aligned} \dot{z}_{21} &= z_{22} \\ \dot{z}_{22} &= (a_1 - a_2 b_1 / b_2) z_{22} \end{aligned} \quad (7)$$

In order for the system to be of minimum phase, the zero dynamics must be stable. In other words, all the roots of the system characteristic equation (7) must be on the left side of the imaginary axis. However, the characteristic equation for this system is:

$$s(s - a_1 + a_2 b_1 / b_2) = 0$$

The two roots of the following equation are both in the origin (considering that $a_1 - a_2 b_1 / b_2 = 0$). Therefore, by choosing an appropriate $v$ in the control law, the system can have stable zero dynamics. Considering the system relative degree value and the system output error as $e_0 = y - y_m$, $v$ is chosen as (8).

$$v = \ddot{y}_m - 2\ddot{e}_0 - 8\dot{e}_0 \quad (8)$$

Additionally, consider that closed-loop poles of the system are arbitrarily chosen as $s = 1 \pm 2.56 j$ [10].

IV. INDIRECT ADAPTIVE FUZZY CONTROLLER DESIGN

According to section II, a portion of input-output formulation of the system that can be linearized is generally presented as (9).

$$\begin{cases} x^{(r)} = f(x, \dot{x}_1, ..., x^{(r-1)}) + g(x, \dot{x}_1, ..., x^{(r-1)}) u \\ y = x \end{cases} \quad (9)$$

In the above equations, $r$ is the relative degree of the system, and $f$ and $g$ are assumed as unknown functions.

In this paper, the control objective is to design a feedback controller $u = u(X, \theta)$ based on the Mamdani fuzzy system, and also to provide an adaption law to regulate the parameter vector $\theta$ such that system output y tracks the desired output $y_m$. $X$ is also a vector representing the system state variables.

$\hat{f}$ and $\hat{g}$ functions are estimations of functions $f$ and $g$ respectively. These approximate functions are attained based on if-then fuzzy rules, and the if-then fuzzy rules are resulted from the system input-output behavior. Since the if-then rules are approximate, some parameters in $\hat{f}$ and $\hat{g}$ are assumed as free parameters. In this way, these parameters can change linearly during the process and help to improve the estimation accuracy as time goes by.

Thus, by assuming vectors $\theta_f$ and $\theta_g$ as the free parameters of $f$ and $g$, $\hat{f}$ and $\hat{g}$ can be stated as follows:

$$\hat{f}(X) = \hat{f}(X, \theta_f)$$
$$\hat{g}(X) = \hat{g}(X, \theta_g)$$

Then, by implementing $\hat{f}$, $\hat{g}$ and the adaption law, the control signal $u$ can be presented as (10).

$$u = \frac{1}{\hat{g}(X, \theta_g)}[-\hat{f}(X, \theta_f) + y_m^{(n)} + K^T e] \quad (10)$$

$$e = y - y_m, \quad e = [e \; \dot{e} \; ... \; e^{(n-1)}]^T$$

Vector $k = [k_n \; ... \; k_1]^T$ is chosen such that all the roots of polynomial (11) are situated in the left side of the imaginary axis.

$$s^n + k_1 s^{n-1} + ... + k_n = 0 \quad (11)$$

Based on the Mamdani fuzzy system and system inputs-outputs attained from the classical controller in off-line

mode, $\hat{f}(X,\theta_f)$ and $\hat{g}(X,\theta_g)$ can be obtained by following these steps:

*Step 1:* For variables $x_i$ ($i = 1, 2, ..., n$) consider $P_i$ number of input fuzzy sets as $A_i^{l_i}$ ($l_i = 1, 2, ..., p_i$). Also, consider $q$ number of output fuzzy sets as $B^l$ ($l = 1, 2, ..., q_j$). These fuzzy sets contain the input-output dynamics behavior of the system.

*Step 2:* Construct the following if-then statements based on the system input-output behavior;

If $x_i$ is $A_i^{l_i}$, then $\hat{f}$ ($\hat{g}$) is $B^{l_{i1}...l_{in}}$;

whereas $1 < i_1, ..., i_n < n$ and $1 < l_{i_1}, ..., l_{i_n} < q$.

There are $\prod_{i=1}^{n} P_i$ number of if-then rules in this process, and the accuracy of these statements is directly related to our knowledge of system input-output behavior.

Now, with the aid of an inference engine, singleton fuzzifier, and centroid defuzzifier, $\hat{f}(X,\theta_f)$ and $\hat{g}(X,\theta_g)$ functions can be represented as follows:

$$\hat{f}(X,\theta_f) = \theta_f^T \xi_f(X)$$
$$\hat{g}(X,\theta_g) = \theta_g^T \xi_g(X)$$

In the above equations, $\xi_f$ ($\xi_g$) are vectors with dimension $\prod_{i=1}^{n} P_i$, and can be stated as follows:

$$\xi_f(X) = \frac{\prod_{i=1}^{n} \mu_{A_i^{l_i}}(x_i)}{\sum_{l_1=1}^{P_1} \cdots \sum_{l_n=1}^{P_n} \left[ \prod_{i=1}^{n} \mu_{A_i^{l_i}}(x_i) \right]}$$

Whereas, by substituting input samples of $f$ or $g$ in $x_i$ s, the vector $\xi_f(X)$ or $\xi_g(X)$ can be specified. Free parameters here are the centers of output fuzzy sets ($\theta_f$ and $\theta_g$ vectors with $\prod_{i=1}^{n} P_i$ dimension). Here, it is assumed that the system has parametric uncertainties, meaning that $f$ and $g$ structures are specified from (3) with unknown parameters. Variables $g(X) = b_2$ and $f(X) = a_2 x_2 + a_3 \sin x_3 + a_4 x_4$ are functions of the state variables $x_2$, $x_3$, and $x_4$, and are considered as inputs to the fuzzy systems. The membership functions represented in Fig. 2 and Fig. 3 are considered as the state variables and system output in their specific domain.

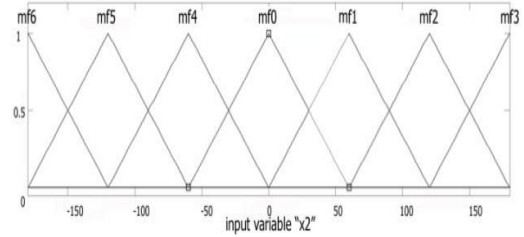

Fig. 2. Membership function for state variable $x_2$ and output signal $x_3$.

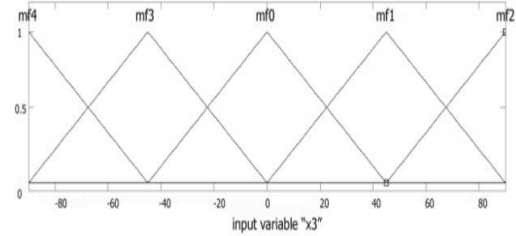

Fig. 3. Membership function for state variables $x_3$ and $x_4$.

Choosing $k = [-0.7 \; 1 \; 10.8 \; 0.7]^T$ for this system, and substitution of this vector in the equation below:

$$u = \frac{1}{g(X)}[-f(X) + y_m^{(n)} + K^T e]$$

The control law will be resulted as (12).

$$u = \frac{1}{b_2}[-f(X) + y_m^{(4)} - 0.7e + \dot{e} + 10.8\ddot{e} + 0.7\dddot{e}] \quad (12)$$

In order to attain $\theta_f$ and $\theta_g$ for adaption law in $f$ and $g$ parameters, Lyapunov function $v$ is chosen as follows:

$$V = \frac{1}{2} e^T P e + e^T P b w + \frac{1}{2\gamma_1}(\theta_f - \theta_f^*)^T(\theta_f - \theta_f^*)$$
$$+ \frac{1}{2\gamma_2}(\theta_g - \theta_g^*)^T(\theta_g - \theta_g^*)$$
$$A^T P + P A = -Q$$

By derivation of the Lyapunov function, the following equation results.

$$\dot{v} = -\frac{1}{2} e^T P e + e^T P b w + \frac{1}{\gamma_1}(\theta_f - \theta_f^*)^T[\dot{\theta}_f + e^T P b \xi_f(X)] + \frac{1}{\gamma_2}(\theta_g - \theta_g^*)^T[\dot{\theta}_g + \gamma_2 e^T P b \xi_g(X) u]$$

Matrix $Q$ is chosen as a diagonal matrix with diagonal values of 1000 and matrix $A$ is chosen as

$$\begin{bmatrix} 0 & 1 & 0 & 0 \\ 0 & 0 & 1 & 0 \\ 0 & 0 & 0 & 1 \\ -0.7 & -10.8 & -1 & 0.7 \end{bmatrix}$$

.Thus, by solving Lyapunov equation, matrix $P$ is attained.

$$P = 10^3 \times \begin{bmatrix} 7.7709 & 0.3740 & -0.5139 & 0.7143 \\ 0.3740 & -4.6545 & -0.9861 & 0.0809 \\ -0.5139 & -0.9861 & 0.2394 & -0.4861 \\ 0.7143 & 0.0809 & -0.4861 & -0.0199 \end{bmatrix}$$

The adaption law must be chosen such that the Lyapunov derivative becomes negative. Therefore, the last two terms of the Lyapunov derivative equation above must be equal to zero.

$$\begin{cases} \dot{\theta}_f = -\gamma_1 e^T Pb \xi_f(X) \\ \dot{\theta}_g = -\gamma_2 e^T Pb \xi_g(X)u \end{cases}, \quad b = [0 \ 0 \ 0 \ 1]^T$$

Also $\gamma_1$ and $\gamma_2$ are chosen as 35 and 6 respectively. Eventually, the attained adaption law is applied in the simulation. So, the general block diagram for the adaptive fuzzy control system is shown in Fig. 4.

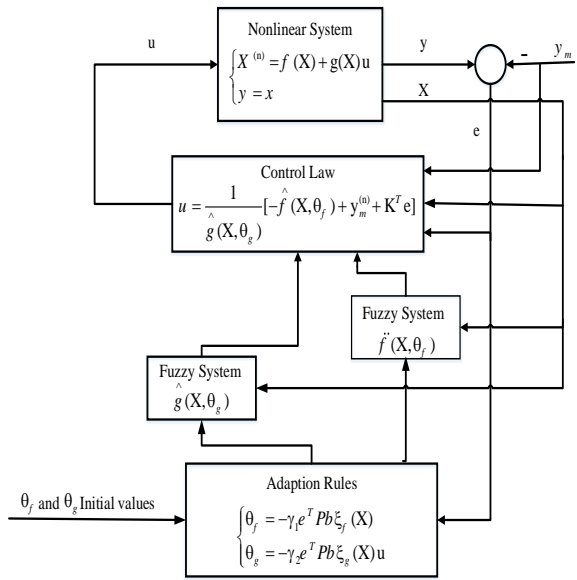

Fig. 4. Block diagram for adaptive fuzzy control system.

## V. SIMULATION RESULTS

The nonlinear rotational inverted pendulum system is simulated with two different feedback controllers; a conventional feedback linearization controller and an adaptive fuzzy controller. The simulation results are shown in this section. Moreover, the output response of the system with adaptive fuzzy controller is analyzed with regard to the existence of parametric uncertainty. The output signal of the mentioned pendulum system is attained by applying a feedback linearization controller as shown in Fig. 5. The desired output signal ($y_m$) is also assumed to be sinusoidal in this simulation. System output response using the adaptive fuzzy controller is presented in Fig. 7, and Figs. 6 and 8 graphs illustrate the tracking error by conventional and proposed controllers respectively.

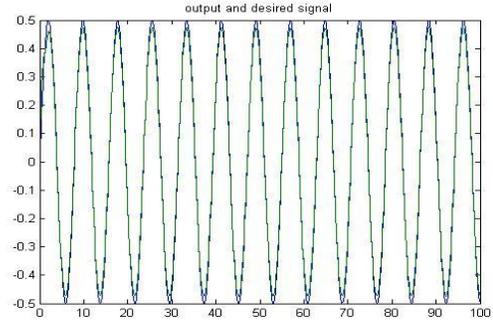

Fig. 5. Desired signal tracking using the classical controller.

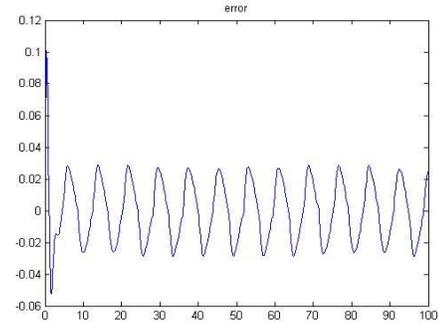

Fig. 6. Tracking error using the classical controller.

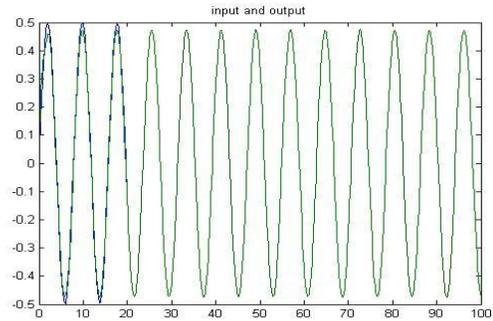

Fig. 7. Desired signal tracking using the adaptive fuzzy controller.

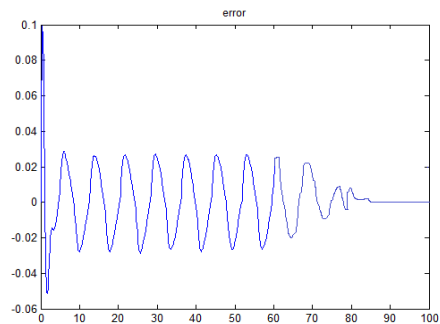

Fig. 8. Tracking error using the adaptive fuzzy controller.

By comparing Figs. 5 and 7, it is proven that reference input tracking results using the adaptive fuzzy controller is more superior to results attained by applying classically designed controller. Tracking error Figs. 6 and 8 also confirm better performance of the proposed controller within this specific system. Indeed, by applying an adaptive fuzzy controller on this system the steady state tracking error is almost zero, however this value is not zero using the classical feedback linearization controller (fluctuates between -0.03 to 0.27).

Additionally, output signal tracking in existence of system parametric changes (uncertainties) is shown in Fig. 9.

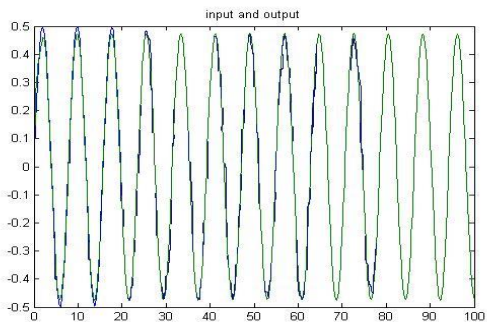

Fig. 9. Desired signal tracking using the adaptive fuzzy controller in existence of system parametric changes.

According to Fig. 9 results, the system with adaptive fuzzy controller is resistant to parametric changes. It can be seen in the graph that the closed loop system using the proposed controller tracks the reference input in the existence of parametric changes, whereas the classical control method shows the appropriate performance only for some fixed values of parameters and it is not reliable in terms of tracking the reference input while the parameters change in the system. Thus, by using the proposed controller there is no need to have advanced knowledge of the system as this controller is the appropriate choice for complicated or unknown systems with parametric uncertainties.

## VI. Conclusion

In this paper, a conventional feedback linearization controller and an adaptive fuzzy controller have been applied to a nonlinear pendulum system, which is a known benchmark control problem, and the results have been analyzed. Comparing the performance of both controllers via the simulation results, the adaptive fuzzy controller has proven to be of higher efficiency and reliability in tracking the reference signal. The proposed adaptive fuzzy controller for this system also has illustrated superior results in existence of system parametric changes (known as a challenging problem in pendulum system control) compared to the conventional feedback linearization controller. For future work, the performance analysis of the adaptive fuzzy controller for this pendulum system in existence of undesired disturbances is recommended.